\documentstyle[prl,aps]{revtex}

\begin{document}
\author{Lev I. Deych,$^\dagger$ D. Zaslavsky,$^\ddagger$ and A.A. Lisyansky$^\ddagger$}
\address{$^\dagger$Department of Physics, Seton Hall University, South Orange, NJ 07079}
\address{$^\ddagger$Department of Physics, Queens College, CUNY, Flushing, NY 11367}
\title{Statistics of the Lyapunov exponent in 1-{\em D} random periodic-on-average
systems\footnote{Phys. Rev. Lett. \textbf{81}, 5390 (1998)}}
\date{\today}
\maketitle

\begin{abstract}
By means of Monte Carlo simulations we show that there are two qualitatively
different modes of localization of classical waves in 1-{\em D} random
periodic-on-average systems. States from pass bands and band edges of the
underlying band structure demonstrate single parameter scaling with
universal behavior. States from the interior of the band gaps do not have
universal behavior and require two parameters to describe their scaling
properties. The transition between these two types of behavior occurs in an
extremely narrow region of frequencies. When the degree of disorder exceeds
a certain critical value the single parameter scaling is restored for an
entire band-gap.

\end{abstract}


In this paper we numerically study localization properties of band gap
states in a one-dimensional periodic-on-average random system (PARS). These
kinds of system were extensively studied in the past in the context of
electron localization, where they were known as Kronig-Penny-like models
(see, for example, Ref. \cite{Erdos,Lifshits,Altshuler}
and references therein). Classical wave versions of 1-{\em D }PARS also
recently attracted a considerable attention \cite
{Bart,Maradudin,Sheng,Maradudin2,Sigalas}. Most of these studies focused
upon localization properties of states from pass (conduction) bands of the
respective initial periodic systems, or states at band edges of the original
spectrum. They were found to behave similarly to the
one-dimensional Anderson model, demonstrating
single-parameter scaling (SPS) and universality \cite{Anderson1}.

Disorder, however, not only localizes states in the conduction bands of 1-%
{\em D} systems, it also gives rise to localized states inside band gaps of
the original spectrum. This is well known in the physics of disordered
semiconductors, where a vast literature on properties of localized states
arising within forbidden gaps of semiconductors exists (see, for example,
book \cite{Efros}). In the case of one-dimensional models, however, these
states have been studied surprisingly little. Particularly, statistical
properties of the Lyapunov exponent (the inverse localization length), $%
\lambda $, for these states have not been studied at all. At the same time,
it turns out that the variance, $var(\lambda )$, of the Lyapunov exponent
contains important information about spectral properties of these systems.
From the frequency dependence of the variance, we find that the band gap
states can be divided into two groups with qualitatively different
localization properties separated by a sharp boundary. This means that
though all states in 1-{\em D} systems are localized, there might be two
qualitatively different regimes of localization. The first regime
corresponds to the band and band edge states, and has regular Anderson
behavior (if disorder is locally weak). The second regime, associated with
the gap states, does not obey SPS and is not universal. The regular
tight-binding Anderson model also demonstrates violation of SPS, when
disorder becomes {\em locally} strong\cite{Shapiro,Stone}. It should be
empasized, therefore, that in our case the absence of SPS is caused not by
the strength of disorder, but by the different nature of the gap states.
Studying how $var(\lambda )$ and the Lyapunov exponent (LE) itself depend
upon the degree of disorder (rms fluctuations of a random parameter, $\sigma 
$) we find that there exists a critical value, $\sigma _{cr}$, at which the
boundary between the groups of states with different localization properties
disappears. At $\sigma >$ $\sigma _{cr}$ all the states have the regular
Anderson-like behavior. It is interesting to note that in this situation SPS
is {\em restored} when disorder becomes stronger, contrary to what one would
expect in the Anderson model.

In the paper we deal with the classical wave version of PARS and consider
localization properties of scalar waves in a 1-$D$ superlattice composed of
two alternating layers $A$ and $B$ with dielectric constants $\varepsilon
_{A}$ and $\varepsilon _{B}$, respectively. The results, however, can also
be applyed to Kronig-Penny-like models of electron localization. We
introduce disorder in the system forcing the thickness $d_{B}$ of the {\em B}
layers to change randomly assuming that $d_{B}$ is drawn independently
from a uniform distribution.

The structure of the described model is periodic on average with the spatial
period equal to $d=d_{A}+\left\langle d_{B}\right\rangle $ and with random
positions of the boundaries between different layers. The states of the
model are characterized by a dimensionless wave number $k=(\omega /c)d$,
where $\omega $ is the frequency and $c$ is the vacuum speed of light.

We study the model by means of the transfer-matrix method. The state of the
system is described by the vector ${\bf u}_{n}$ with components representing
the wave field, $E_{n},$ and its spatial derivative, $E_{n}^{\prime }$. The
evolution of the vector ${\bf u}_{n}$ is controlled by the matrix ${\bf X}%
_{n}$: ${\bf u}_{n+1}={\bf X}_{n}{\bf u}_{n}$, where ${\bf X}_{n}$ is
determined as follows:

\begin{equation}
{\bf X}_{n}=\left( 
\begin{array}{cc}
\cos (k_{n}d_{n}) & \displaystyle{\frac{1}{k_{n}}}\sin (k_{n}d_{n}) \\ 
-k_{n}\sin (k_{n}d_{n}) & \cos (k_{n}d_{n})
\end{array}
\right) ,  \label{Tmatrix}
\end{equation}
where $k_{n}=k\sqrt{\epsilon _{n}}$. LE can be computed in accordance with
the following definition: 
\begin{equation}
\lambda =\lim_{L\rightarrow \infty }\frac{1}{L}\left\langle \ln \left( {%
\frac{\parallel {\bf X}^{(N)}{\bf u}_{0}\parallel }{\parallel {\bf u}%
_{0}\parallel }}\right) \right\rangle ,  \label{Lyapunov}
\end{equation}
where $L$ is the length of the system, the matrix ${\bf X}^{(N)}$ is a
product of all ${\bf X}$-matrices corresponding to each layer, ${\bf X}%
^{(N)}=\prod_{1}^{N}{\bf X}_{n}$, and ${\bf u}_{0}$ is a generic vector. LE
determined according to Eq. (\ref{Lyapunov}) is a self-averaging quantity.
For finite systems, however, it exhibits fluctuations that are the main
object of study of this paper.

In computer simulations we generated a sequence of random ${\bf X}$-matrices
in accord with the model described above. The parameters of the model were chosen as follows: $\varepsilon
_{A}=1$, $\varepsilon _{B}=1.2$; $d_{A}=1$ and the mean width of the $B$ layers $\left\langle d_{B}\right\rangle =1$. The standard deviation $\sigma$ of $d_B$ is a measure of the disorder in the model. 
The results of the calculations are presented in figures below. Fig. 1 shows
the frequency dependence of LE and its variance, $var(\lambda )=\langle
\lambda ^{2}\rangle -\langle \lambda \rangle ^{2}$, for frequencies covering
one of the band-gaps. The behavior of $\lambda $ coincides with the results
reported previously in Ref. \cite{Bart,Maradudin,Sigalas}. The variance of $%
\lambda $ at the same time demonstrates anomalous non-monotonic behavior
with two maxima inside the band. In the region between the maxima, $%
var(\lambda )$ changes {\em oppositely} to $\lambda ,$ approaching its
minimum value at the center of the gap. Such a behavior is clearly
inconsistent with SPS, which dictates that $var(\lambda )=(2/L)\lambda $\cite
{Anderson1,Shapiro,Stone2}. When the dispersion of our random variable
increases, the maxima of $var(\lambda )$ moves toward the center of the gap,
and the central minimum raises. At some critical value, $\sigma _{cr}$,
double-peaked structure of $var(\lambda )$ dissapear, and at $\sigma >\sigma
_{cr}$ the variance exhibits only one maximum and goes along with LE itself.
The results presented in this figure show that localization properties of
band and band-edge states are different from those of the gap states. This
difference sharply manifests itself in the dependence of LE upon $\sigma $
shown in Fig. 2 for different $k$. This dependence undergoes a strong
qualitative change when the wave number moves toward the center of the gap.
The difference in behavior of \ LE corresponding to the states from the pass
band and the gap was previously discussed in Refs. \cite
{Sheng,Maradudin2,Zaslavsky}. Our results show that the change between
different shapes of the function $\lambda (\sigma )$ occurs within a  narrow
frequency interval of the order of magnitude of $0.01\%$ suggesting
existence of a well defined boundary between different groups of state. It
is remarkable that this boundary coincides with the position of the
respective maximum of $var(\lambda )$. For $\sigma >$ $\sigma _{cr}$, where $%
\sigma _{cr}$ corresponds to the transformation between different types of
behavior of $var(\lambda )$, all functions $\lambda (\sigma )$ for different 
$k$ merge together demonstrating behavior independent of frequency of the
states.

Fig. 3 presents $var(\lambda )$ directly plotted versus $\lambda $. In order
to obtain this figure, we combine frequency dependences of LE and its
variance in the region covering several bands of the parent periodic system.
This figure illustrates the violation of the SPS in our system. It is clear
from comparison between this plot and the plot in Fig. 1 that anomalous
non-linear dependence of $\ var(\lambda )$ versus $\lambda $ is caused by
the states from the inner region of the band gap between the two maxima of $%
\ var(\lambda )$. The multibranch structure of the plot shows the lack of
universality in this dependence: it is different for different band gaps.
Indeed, we checked that different branches originate from frequencies
corresponding to different band-gaps. Double-lines, which form each of the
branches, correspond to different halves of the same band gap. The insert in
Fig. 3 presents one of the branches that was obtained by filtering out all
the frequencies except those that belong to one half of a band-gap.

The nonuniversal behavior of states from different band-gaps and a violation
of SPS are closely related phenomena. The intimate relationship between
nonuniversality and SPS is demonstrated in Fig. 4. These plots show the
evolution of the dependence of $var(\lambda )$ versus $\lambda $ with an
increase of the value of the standard deviation, $\sigma ,$ of the layers'
thickness. One can observe that with an increase of $\sigma $, the variance
of $\lambda $ approaches a regular linear dependence upon $\lambda $. And
along with this, all the branches from the different band-gaps merge into
one universal curve. It happens at approximately the same values of $\sigma $
at which the maxima on the frequency dependence of $var(\lambda )$
disappear, signaling about complete eroding of the initial band structure by
the disorder. However, a destruction of the initial spectrum is not a
uniform (in terms of frequencies) process - some bands disappear at smaller
disorder than others. Therefore, different branches of the graph in Fig. 4
return to the universal behavior at different values of the disorder
parameter, $\sigma $.

The presented results suggest that the probability distribution function of
LE, $P(\lambda )$, in the case of the inner gap states, cannot be described
by a single parameter. We find, however, that in the asimptotic limit $%
L\rightarrow \infty $, $P(\lambda )$ retains its gaussian form with $%
var(\lambda )\propto 1/L$. That means that statistical properties of the gap
states are characterized by two independent parameters. The similar
situation occurs in the Anderson model in the case of locally strong
scattering \cite{Shapiro,Stone}. In our situation the local scattering,
characterized by the mismatch between the dielectric parameters of the
layers, is weak. The considered system, therefore, presents an example when
two-parameter scaling occurs due to properties of the initial spectrum of
the system rather then due to the strength of the disorder.

To suggest a qualitative explanation of the results let us first consider
the nonmonotonic behavior of $var(\lambda )$. It is more convenient to start
with the states arising in the center of the forbidden band, which exhibit
the smallest fluctuations of $\lambda $. The reason for this behavior lies
in the fact that states deep in the forbidden band can only arise due to a
strong deviation from the initial periodic structure. The probability of
such events is small, and such defect configurations, occurring at different
parts of the structure, are separated from each other by distances much
greater than localization lengths of states emerging due to these
deviations. The wave functions corresponding to different states do not
overlap and such configurations can be considered as isolated defects. It is
well known (see, for example, \cite{Lifshits}) that the localization length
in this case is determined solely by the initial spectrum of the system and
the frequency of the corresponding local state. Therefore, fluctuations of $%
\lambda $ would be equal to zero in the idealized situation of isolated
defects. In the real case, we have small fluctuations due to overlap of
exponential tails of different states, and rare occasions when two defects
appears close to each other.

When $k$ moves from the center of the band-gap toward the boundary, the
density of the configurations responsible for the states at corresponding
frequencies and their localization lengths increase. First this leads to an
increase in the fluctuation of LE because the localization length is
determined now by complexes of several interacting defect configurations,
and becomes sensitive to the fluctuating structure of such complexes.
However, as the band boundary is approached, the effect of self-averaging
comes into play. With an increase of spatial overlap between states
localized at different centers, self-averaging of LE becomes more effective,
leading, therefore, to the decrease of $var(\lambda )$. The interplay
between these two tendencies generates nonmonotonic behavior of $var(\lambda
)$ presented in Fig. 1.

Arguments similar to that presented above can explain the difference in the
the $\sigma $-dependence of LE for states from the inner part of the
band-gaps and boundary states. The behavior of LE for states from the
pass-bands and states at the band edges is well understood: $\lambda \propto
\sigma ^{2}$ for states from the pass bands, and $\lambda \propto \sigma
^{2/3}$ at the band edges \cite{Bart}. It is also clear that when $\sigma
\rightarrow 0,$ LE for states from the band-gap tends to a non zero value,
which is just an inverse penetration length through the forbidden band. This
penetration length obviously decreases when frequency approaches the center
of the band-gap. Disorder results in emergence of a small number of states
with frequencies in the inner part of the gap. These states enhance
propagation of waves due to a process similar to resonant tunneling,
reducing by this means LE. These arguments cannot be applied to states at
the band edge because their penetration length at $\sigma =0$ is so large
that even small disorder makes them indistinguishable from the states at the
pass-band side of the edge.

It is clear from the presented results that the violation of SPS and the
absence of universality in the considered system are due to nonmonotonic
behavior of $var(\lambda )$. It is interesting, however, to examine this
behavior from the standpoint of general arguments leading to SPS \cite
{Anderson1}. According to these arguments, SPS is realized when the
localization length $l_{loc}=\lambda ^{-1}$ is much greater than the phase
randomization length $l_{ph}$. In the random phase model \cite{Anderson1} it
is assumed that phase randomization occurs at a microscopic scale, and,
violation of the inequality $l_{loc}\gg l_{ph}$ means that the localization
length also becomes of microscopic size. It is natural, therefore, not to
expect universality in this case. In our situation, however, localization
length remains macroscopical even for states deep inside the gap. Therefore,
the absence of SPS is not related to the influence of microscopic details of
the system. At the same time one could expect the increase in $l_{ph}$ in
this spectral region, because the phase of a wave changes only when it
crosses a localized state, which are very rare in this spectral region. One
can assume, therefore, that for the state from the inner region of the band
gap one can have $l_{ph}>l_{loc}$ even though both these lengths remain {\em %
macroscopical}. In order to check this assumption, we studied a distribution
of the phase, and determined the phase randomization length using simple
Neyman-Person $\chi ^{2}$ statistical tests to check the uniformity of the
distribution. We used the plane wave representation for the transfer-matrix,
and following Ref. \cite{Stone}, define the phase, as a relative phase
responsible for the scaling behavior of LE. The results of the
simulation confirm that the phase randomization length increases inside the
gap, and exceeds the localization length for the states between maxima of $%
var(\lambda )$. We also found that the transition from the inequality $%
l_{ph}<l_{loc}$ to $l_{ph}>l_{loc}$ occurs approximately in the same
critical region, which separates the band-edge states from the inner states
in Figs.\thinspace 1 and 2.

To conclude, Monte-Carlo computer simulation of wave localization properties
of a one-dimensional random periodic-on-average system shows that the
spectrum of the system can be divided into two groups of states with
qualitatively different statistical behavior of the Lyapunov exponent and a
narrow critical region separating these groups. The first group combines
states from pass-bands of the parent periodic system and band edge states
from the band-gaps. The statistical properties of this group are similar to
the properties of the Anderson model and can be described by the
single-parameter-scaling. The other group consists of states from the inner
region of the band-gaps and demonstrate a number of anomalies. The variance
of LE, for this group decreases when the frequency approaches the center of
a band-gap resulting in a strong deviation from
single-parameter-scaling-like behavior. The dependence of $var(\lambda )$
upon LE is nonuniversal for these states and is different for different band
gaps. The transition between the two groups is also accompanied by the
reversal of the relationship between the localization length and phase
randomization length: for a pass-band and band edge states $l_{ph}<l_{loc}$,
and $l_{ph}>l_{loc}$\ for inner band-gap states. An increase in the degree
of randomness beyond a well defined critical value erodes the difference
between the groups and returns the system to the regular
single-parameter-scaling mode of behavior. This can be interpreted as a
complete destruction of the original band structure by disorder.

We are pleased to acknowledge stimulating discussions with B.L. Altshuler. We wish to thank S. Schwarz for reading and commenting
on the manuscript. This work was supported by the NSF under grant No.
DMR-9632789 and by a CUNY collaborative grant.

\clearpage

\section*{Figure Captions}

\noindent Fig. 1. Frequency dependence of the Lyapunov exponent, $\lambda ,$
and its variance, $var(\lambda ),$ for the frequencies covering a band-gap ($%
2.83<k<2.88$) and band edges of the underlying structure. The graph was
obtained for a system with $200$ layers, so that states with $2.82<k<2.89$
have localization lengths shorter than the system's size.

\noindent Fig. 2. The dependence of the Lyapunov exponent, $\lambda ,$ upon
the standard deviation, $\sigma ,$ of the random layers' thicknesses for
different $k$ in the vicinity of $k=2.835$, which approximately corresponds
to the left maximum of $var(\lambda )$ shown in Fig. 1.

\noindent Fig. 3. Variance of $\lambda $ versus $\lambda $. The graph was
obtained by means of combining frequency dependencies of $var(\lambda )$ and 
$\lambda $ for the frequency region covering four band-gaps of the parent
periodic structure. The insert corresponds to frequencies from the left half
of the second band-gap.

\noindent Fig. 4. Evolution of the graph from Fig. 3 with increase of $%
\sigma $. It is seen how different branches raise gradually and form the
universal curve corresponding to the prediction of SPS. Note that $\sigma
=0.08$ on the second graph is close to, but still smaller than, the critical
value of $\sigma $ that marks the start of the universal behavior of the
graphs in Fig. 2.

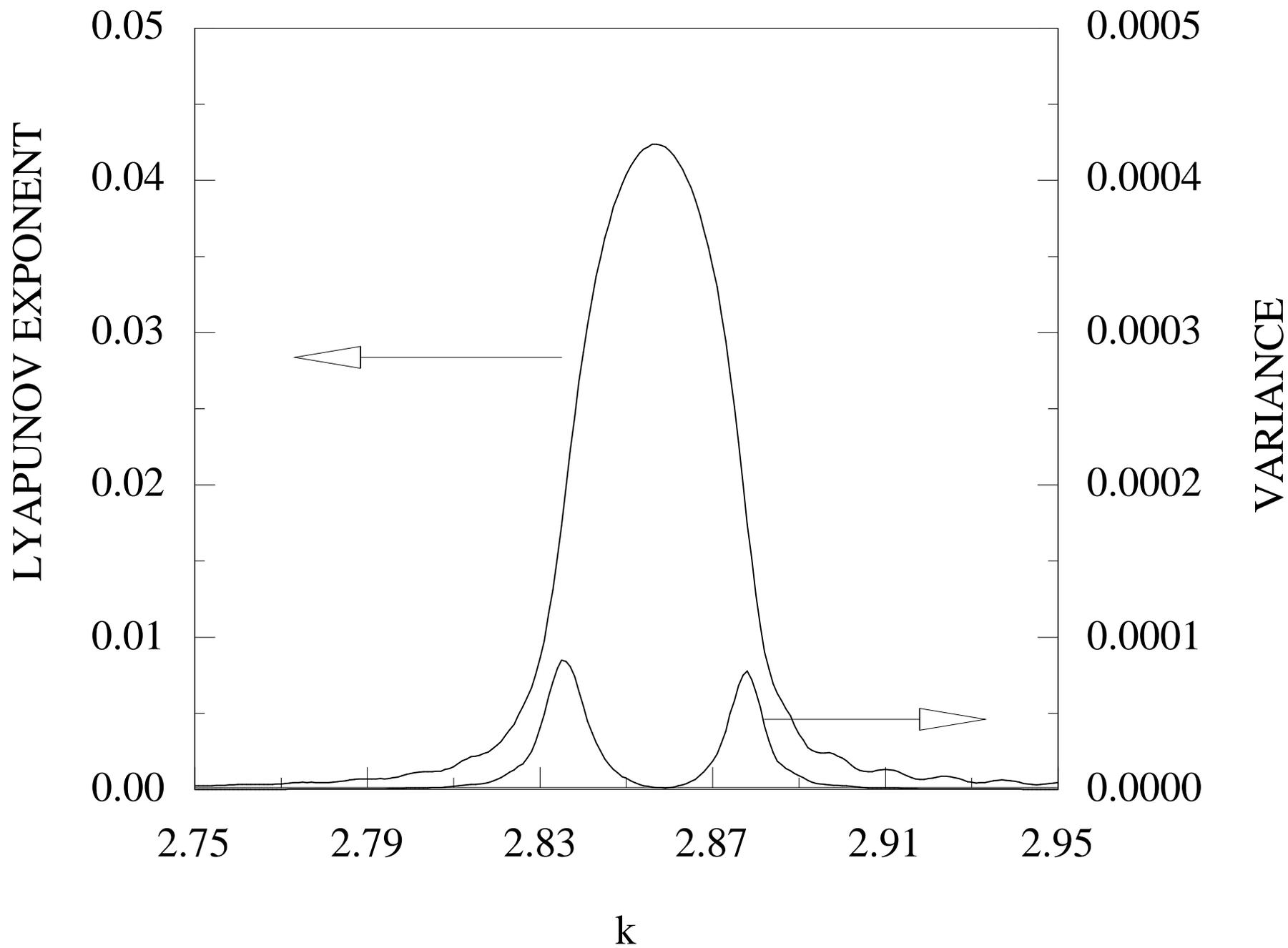
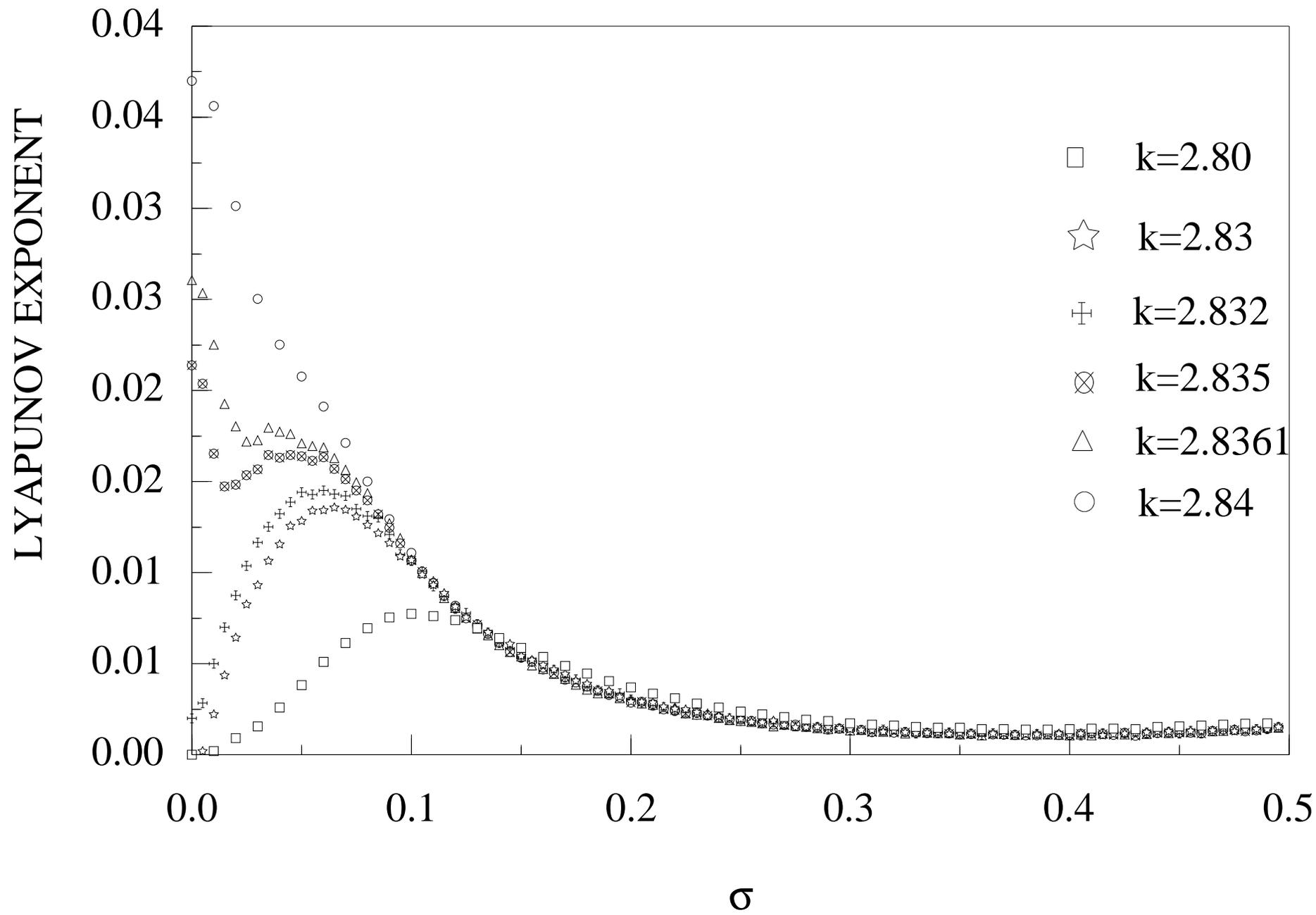
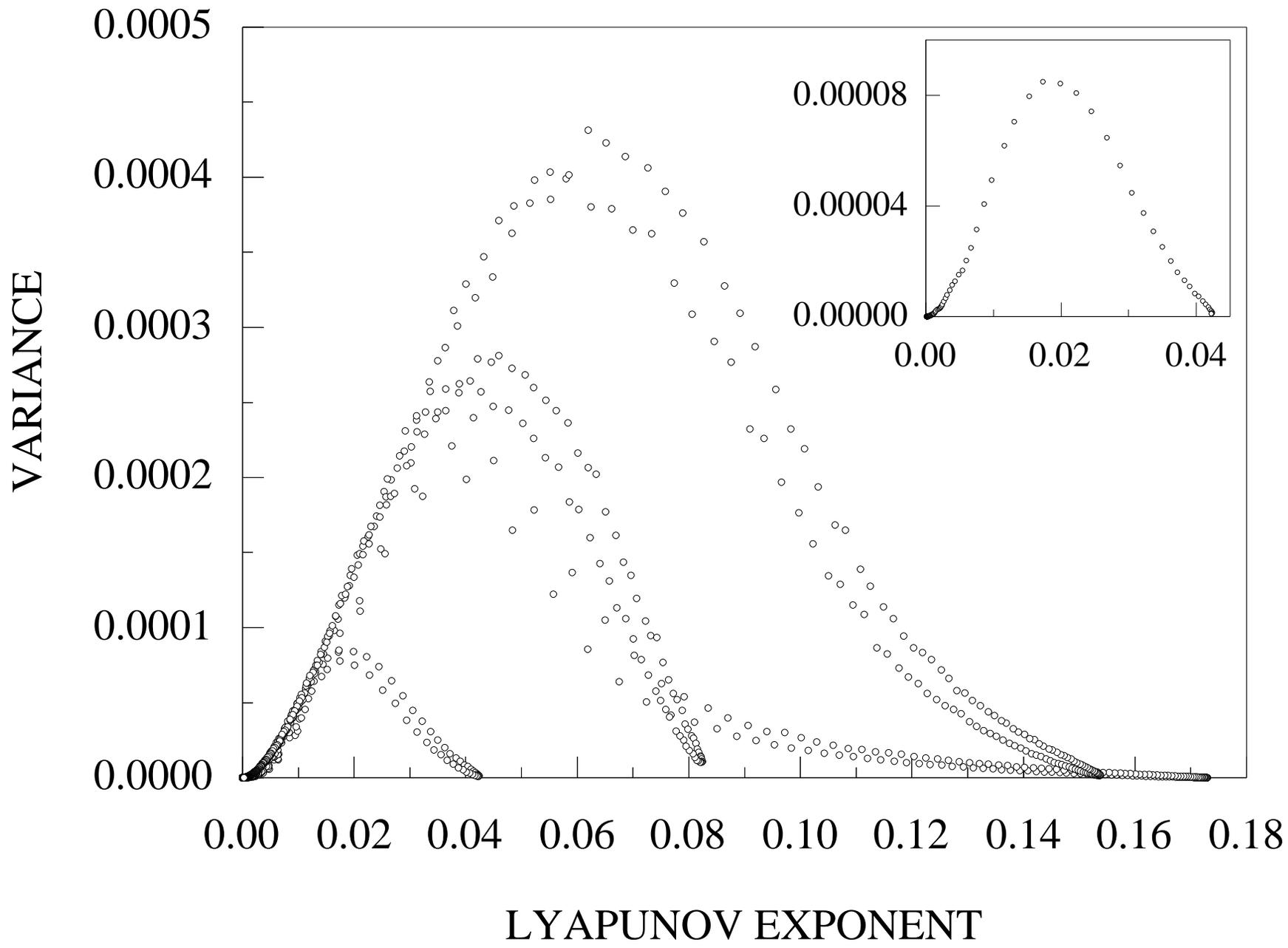
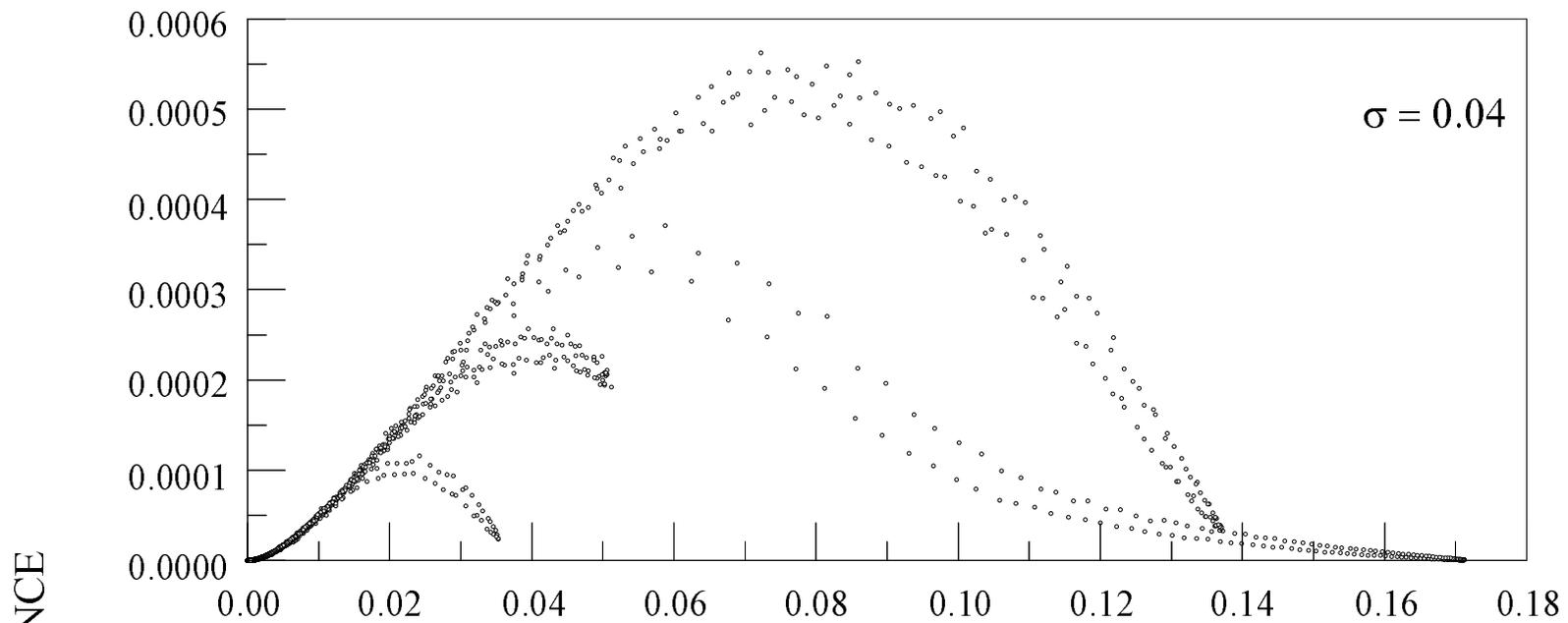
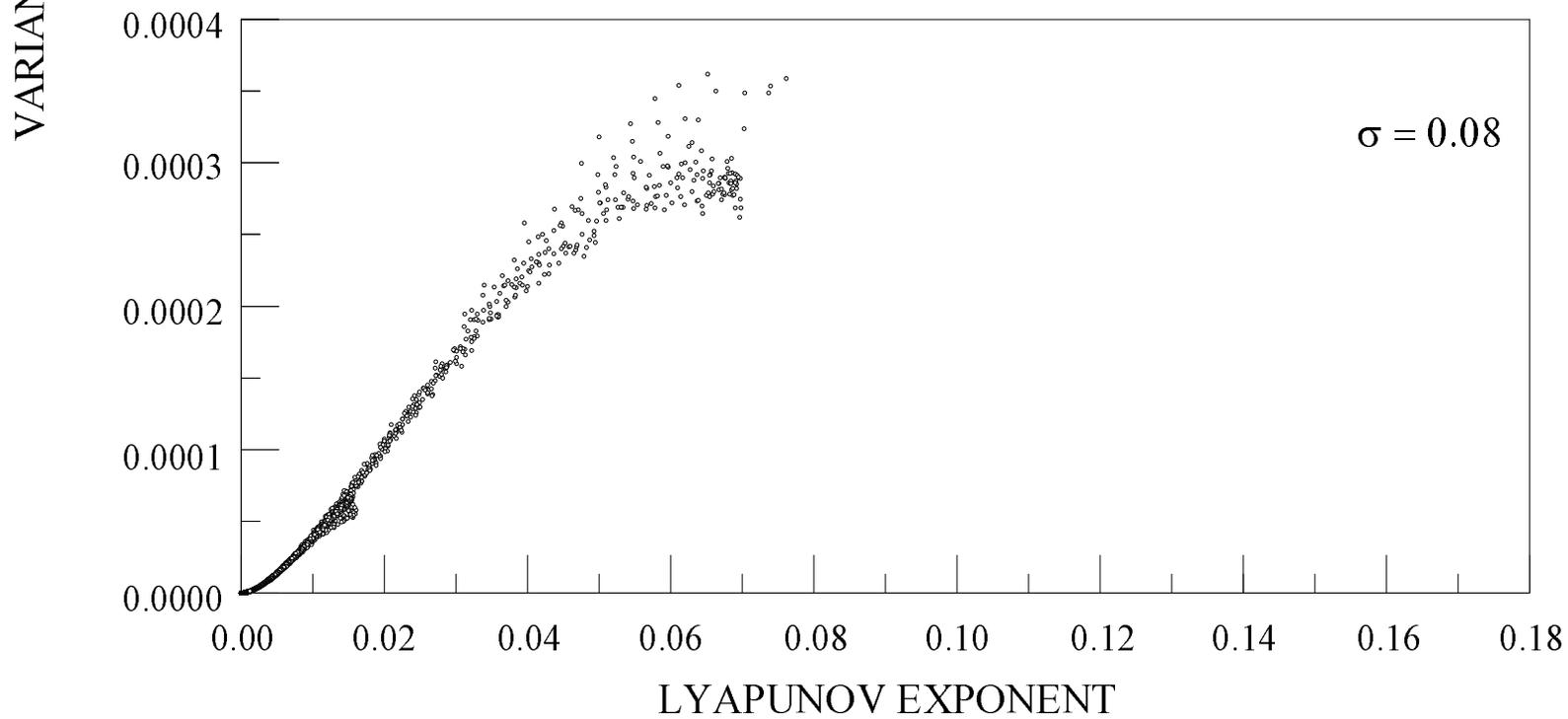
\end{document}